\documentclass[conference]{IEEEtran}
\IEEEoverridecommandlockouts

\usepackage{cite}
\usepackage{amsmath,amssymb,amsfonts}
\usepackage{mathrsfs}
\usepackage{algorithm2e}
\usepackage{algpseudocode}
\usepackage{textcomp}
\usepackage[utf8]{inputenc}
\usepackage[english]{babel}
\usepackage[pdftex]{graphicx}
\usepackage{pgfplots}
\pgfplotsset{compat=1.17}
\usepackage{pgfplotstable}
\usepackage{filecontents}
\usepackage{subcaption}
\usepackage{eso-pic}

\newcommand{\lSec}[1]{\label{sec:#1}}
\newcommand{\rSec}[1]{Section \ref{sec:#1}}

\usepackage{xcolor}
\def\BibTeX{{\rm B\kern-.05em{\sc i\kern-.025em b}\kern-.08em
    T\kern-.1667em\lower.7ex\hbox{E}\kern-.125emX}}
\begin{document}

\newcommand\AtPageUpperMycenter[1]{\AtPageUpperLeft{%
 \put(\LenToUnit{0.13\paperwidth},\LenToUnit{-1cm}){%
     \parbox{0.8\textwidth}{\raggedleft\fontsize{9}{11}\selectfont #1}}%
 }}%
\newcommand{\conf}[1]{%
\AddToShipoutPictureBG*{%
\AtPageUpperMycenter{#1}
}
}

\newcommand{\system}{SmartSlice}

\title{\system: Dynamic, self-optimization of application’s QoS requests to 5G networks
{\footnotesize \textsuperscript{}}
\thanks{\textsuperscript{*} Work done as an intern at NEC Laboratories America, Inc.}
}




\author{\IEEEauthorblockN{Kunal Rao, Murugan Sankaradas, Vivek Aswal\textsuperscript{*} and Srimat Chakradhar}
\IEEEauthorblockA{
\textit{Integrated Systems}\\
\textit{NEC Laboratories America, Inc.}\\
Princeton, NJ\\
\{kunal,murugs,vaswal,chak\}@nec-labs.com}
}

\maketitle
\conf{IEEE Software Defined Systems (SDS), Fog and Mobile Edge Computing (FMEC); SDS-FMEC 2021}

\begin{abstract}

Applications can tailor a network slice by specifying a variety of QoS attributes related to application-specific performance, function or operation.
However, some QoS attributes like guaranteed bandwidth required by the application do vary over time. For example, network bandwidth needs of video streams from surveillance cameras can vary a lot depending on the environmental conditions and the content in the video streams. 
In this paper, we propose a novel, dynamic QoS attribute prediction technique that assists any application to make optimal resource reservation requests at all times. Standard forecasting using traditional cost functions like MAE, MSE, RMSE, MDA, etc. don't work well because they do not take into account the direction (whether the forecasting of resources is more or less than needed), magnitude (by how much the forecast deviates, and in which direction), or frequency (how many times the forecast deviates from actual needs, and in which direction). The direction, magnitude and frequency have a direct impact on the application's accuracy of insights, and the operational costs. We propose a new, parameterized cost function that takes into account all three of them, and guides the design of a new prediction technique. To the best of our knowledge, this is the first work that considers time-varying application requirements and dynamically adjusts slice QoS requests to 5G networks in order to ensure a balance between application's  accuracy and operational costs. In a real-world deployment of a surveillance video analytics application over 17 cameras, we show that our technique outperforms other traditional forecasting methods, and it saves 34\% of network bandwidth (over a $\sim$24 hour period) when compared to a static, one-time reservation. 

\end{abstract}


\section{Introduction}
\lSec{introduction}
Smart sensors (or devices) sense and produce raw data streams that are being used for various types of analytics. These sensors or ``things" are connected to the Internet, giving rise to a new paradigm called Internet of Things (IoT). The total number of IoT devices in the world is estimated to grow to 41.6 billion by the year 2025 and the amount of data generated from these devices is expected to grow to 79.4 zettabytes by 2025 \cite{iot-link}. All this data from so many IoT devices is typically transmitted  over 5G networks to a remote location for analytics processing. 

\begin{figure}[ht]
\centering
\includegraphics[width=0.99\linewidth]{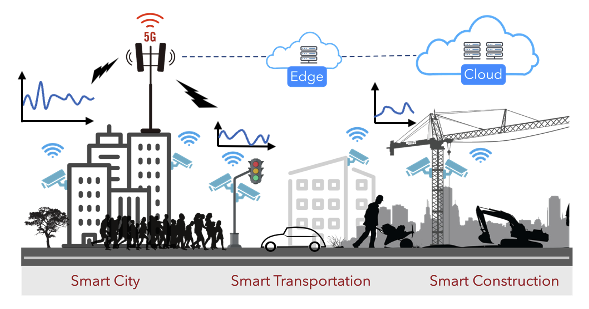}
\caption{Applications on 5G networks}
\label{fig:deployment}
\end{figure}

With the advent and growth of 5G, the rate at which these IoT devices are being installed and deployed has suddenly accelerated. According to Ericsson, by the year 2025, they predict around 5 billion IoT devices will be connected over cellular network \cite{ericsson-link}. The promise of 5G to pack over 1 million IoT devices in 1 square kilometer (massive Machine Type Communication - ``mMTC"), coupled with Ultra Reliable Low Latency (``uRLL") and Enhanced Mobile Broadband (``eMBB"), is driving the deployment of numerous IoT sensors and design of new applications, which were not possible before. For example, consider Intelligent Transportation Systems (ITS), which is a new and emerging application. It is leveraging 5G and IoT sensors to improve various aspects in transportation, including autonomous driving, collision avoidance, traffic planning, etc. Several AR/VR (augmented reality, virtual reality) applications are also emerging, wherein high bandwidth video data transmission and low latency response times are possible due to 5G. Such a scenario of emerging, new applications in a city-scale deployment of sensors (with data being transmitted over 5G for remote processing) is shown in Fig. \ref{fig:deployment}.


\begin{figure*}[]
\begin{subfigure}[]{0.5\textwidth}
\centering
    \includegraphics[width=0.9\textwidth]{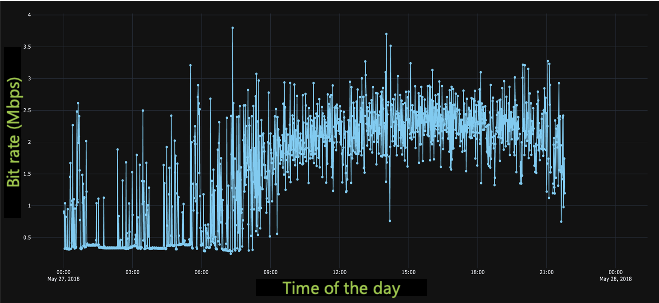}
    \caption{Camera 1}
\end{subfigure}%
\begin{subfigure}[]{0.5\textwidth}
\centering
    \includegraphics[width=0.9\textwidth]{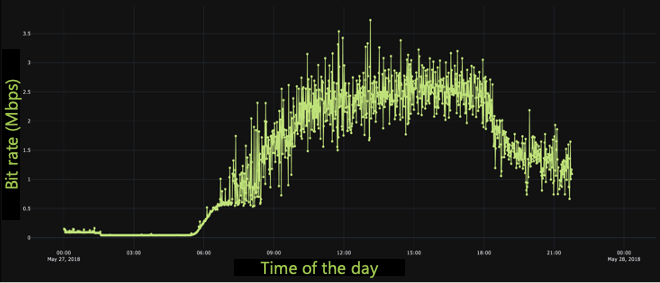}
    \caption{Camera 2}
\end{subfigure}
\begin{subfigure}[]{0.5\textwidth}
\centering
    \includegraphics[width=0.9\textwidth]{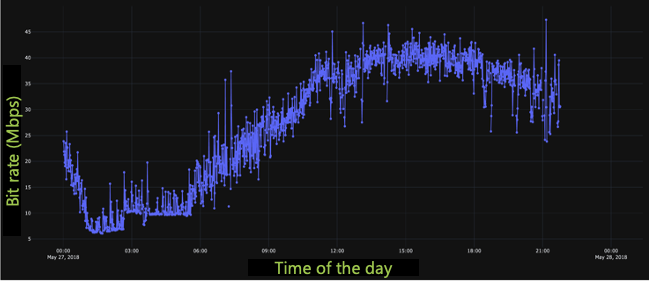}
    \caption{Camera 3}
\end{subfigure}%
\begin{subfigure}[]{0.5\textwidth}
\centering
    \includegraphics[width=0.9\textwidth]{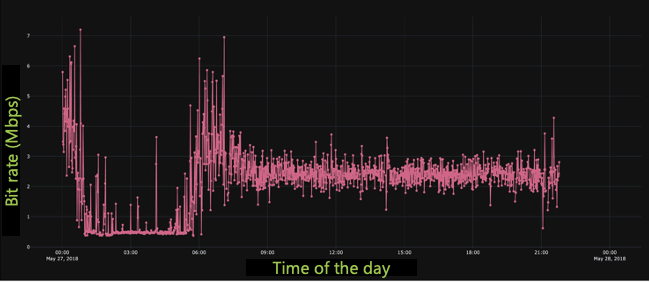}
    \caption{Camera 4}
\end{subfigure}
\caption{Network traffic from surveillance cameras (1 day)}
\label{network-traffic-profiles}
\end{figure*}


In most applications, the network bandwidth required for transporting the raw data streams produced by sensors is usually time-varying because the amount of sensor data changes depending on the environmental conditions and content in the data streams. For example, in case of video surveillance application, the bit rate of the video data stream changes depending on the observed scene, as shown across different cameras in Fig. \ref{network-traffic-profiles} (from a real deployment). If there is no change or relatively less change from one frame to another, then the bit rate drops, whereas, if there is quite a lot of activity and significant changes occur between frames, then the bit rate increases substantially. This is an artifact of the compression algorithms used within the video cameras, and the amount of data produced by the camera varies significantly over a period of time.

5G networks guarantee Quality of Service (QoS) to applications through network slicing. Network slicing sets up end-to-end logical/virtual networks over a common physical communication infrastructure. Applications can request specific QoS by specifying appropriate network slice attributes like the latency, throughput, bandwidth (guaranteed and maximum), packet error rate, reliability, duration of slice, etc. One such abstraction for network slice specification is provided by AppSlice \cite{appslice-nof}. Once a slice is created for an application, this network slice is isolated from other slices, since the network slice has dedicated virtual network resources. Applications are charged based on the amount of network resources they request/reserve, and not on the basis of actual amount of network resources used by an application \cite{5gaa}. 
Thus, it is in the best interest of the application to request only resources that it will actually use (to save operational costs). Desired behavior for any application is that there should be \textit{minimum or no oversubscription} i.e. actual network bandwidth need should not exceed the reserved bandwidth; there should also be \textit{minimum or no undersubscription} i.e. bandwidth used by the application is the same or close to the reserved bandwidth to avoid paying for unused but reserved network bandwidth. As an added incentive, if the application requests fewer network resources, then the chances of 5G network granting the request are much higher.

Our key contributions in this paper are:

\begin{enumerate}
\item We propose a novel QoS request prediction technique, specifically, network bandwidth prediction technique, which any analytics application can use to dynamically self-optimize application's QoS requests to 5G networks
\item We propose a new cost function, which takes into account the magnitude, direction and frequency of the oversubscription or undersubscription, and enables the design of a new prediction technique that continuously optimizes the network bandwidth requests to 5G networks.
\item We show that in a real-world deployment of video analytics application over 17 cameras, our adaptive prediction technique outperforms other traditional forecasting methods, and it is able to save 34\% of network bandwidth (over a 24-hour period) when compared to a one-time static bandwidth request at the time of creation of the slice. 

\end{enumerate}

The rest of the paper is organized as follows. We discuss the related work in \rSec{related}. In \rSec{motivation-system-overview}, we motivate the need for augmenting applications with additional intelligence to modulate QoS requests to 5G networks. In \rSec{system-overview}, we present a system-level overview of how our technique is used by analytics applications in real-world deployments. We discuss our new cost function in \rSec{cost-function} and present our novel prediction technique in \rSec{prediction-technique}. Later, in \rSec{experiments-results} we describe our experiments and the results we obtained by using our proposed technique in a real-world surveillance deployment. Finally, we conclude in \rSec{conclusion}.

\section{Related work}
\label{sec:related}
Various forecasting and resource allocation techniques have been proposed to assist the network operator to optimize network slicing in 5G networks\cite{NECLE-NSB} \cite{NECLE-RLNSB} \cite{Bio-Resource-Allocation} \cite{dynamic-reservation-rl}. These techniques are tailored towards 5G network infrastructure providers/operators, and they discuss ways to efficiently handle slicing requests from multiple applications (multi-tenancy). Some techniques focus on a single application within a single network slice, and they use machine learning techniques to forecast future capacity demands\cite{deepcog}. However, they also present their work from a 5G network operators point of view. In contrast, our proposal is application-centric rather than infrastructure or 5G network operator-centric. Our technique can be used by any application to dynamically self-adapt the network slice QoS requests over a variety of carrier or private 5G network operators. 

Several network traffic prediction techniques have been proposed~\cite{joshi2015review}: linear time series models (ARMA \cite{arma}, ARIMA \cite{arima}, SARIMA \cite{sarima}, AARIMA \cite{aarima}, EPTS \cite{epts}), nonlinear time series models (GARCH \cite{garch}, Neural Network techniques \cite{nn-1} \cite{nn-2} \cite{nn-3}), hybrid models (combination of linear and non-linear models e.g. ARIMA + GARCH \cite{arima-garch}) and decomposed model \cite{decomposed-model}, where time series is decomposed into multiple components. There are also models and prediction techniques based on decision tree regression \cite{decision-tree} and support vector regression \cite{svr}. These prediction models use traditional error metrics to evaluate prediction accuracy, and these metrics include Mean Absolute Error (MAE), Mean Square Error (MSE), Root Mean Square Error (RMSE), Normalized Root Mean Square Error (NRMSE), Mean Percentage Error (MPE) and Mean Absolute Percentage Error (MAPE). These metrics mostly focus on absolute error and optimization techniques try to keep the error as low as possible, without considering the direction or frequency of the error. This may be okay for some scenarios, but in our scenario the direction as well as the frequency of error matters. This was also pointed out in \cite{deepcog} \cite{alternative-measure-of-forecasting-accuracy}. We cannot use the traditional error metrics and we define a new cost function to capture the tradeoff between accuracy of insights and operational costs. We use the cost function to guide us in the design of a new prediction technique. Unlike prior prediction techniques, our prediction technique prioritizes reduction in magnitude and frequency of oversubscription (direction), which is more critical than undersubscription in video analytics applications.
\section{Motivation}
\lSec{motivation-system-overview}
\begin{figure}[ht]
\centering
\includegraphics[width=0.99\linewidth]{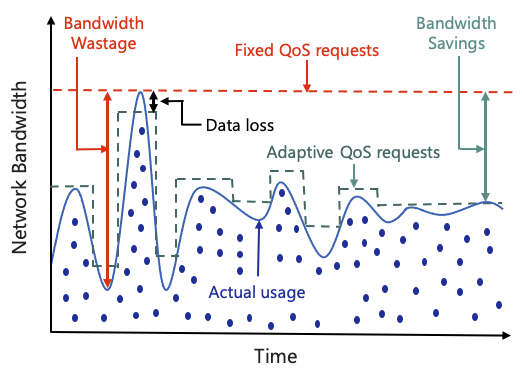}
\caption{Adaptive QoS requests}
\label{fig:adaptive}
\end{figure}

As mentioned in \rSec{introduction}, it is desirable for the application to have \textit{minimum or no oversubscription} and \textit{minimum or no undersubscription} to avoid any data loss (which can result in loss of accuracy of insights from the application), or non-usage of guaranteed network resources. Without knowing the network traffic profile and the usage pattern, an obvious method to issue QoS requests to 5G networks in order to prioritize accuracy over cost, is to request the worst case required network bandwidth. This is shown in Fig. \ref{fig:adaptive} as ``Fixed QoS requests". In response to such a request, the 5G network will grant the worst case required network bandwidth, if available, and then the application may choose to use it however it wants. Now, the actual network traffic usage varies over time (this is shown as ``actual usage"), and there can be surplus network bandwidth reserved. The surplus results in additional operational costs, which can be avoided. Now, in order to reduce the wastage, the intuition is that instead of issuing static, one-time, worst case required network bandwidth request to the 5G network, it may be beneficial to \textit{periodically} send QoS requests to the 5G network, each time \textit{adapting} and \textit{predicting} the required network bandwidth and then issuing QoS request with the predicted network bandwidth. Dynamic adaptation of QoS requests is shown by ``adaptive QoS requests" in Fig. \ref{fig:adaptive}. Such modulation of QoS requests can lead to significant savings in network bandwidth, as shown in the figure as ``Bandwidth savings". However, occasionally, if the prediction is less than the network bandwidth actually required, then there could be some data that won't be transmitted. This is shown in the figure as ``Data Loss". Thus, ``adaptive QoS requests" have the potential for substantial network bandwidth savings, and significant reduction of overall operational costs for the application.

\begin{figure*}[ht]
\centering
\includegraphics[width=0.84\linewidth]{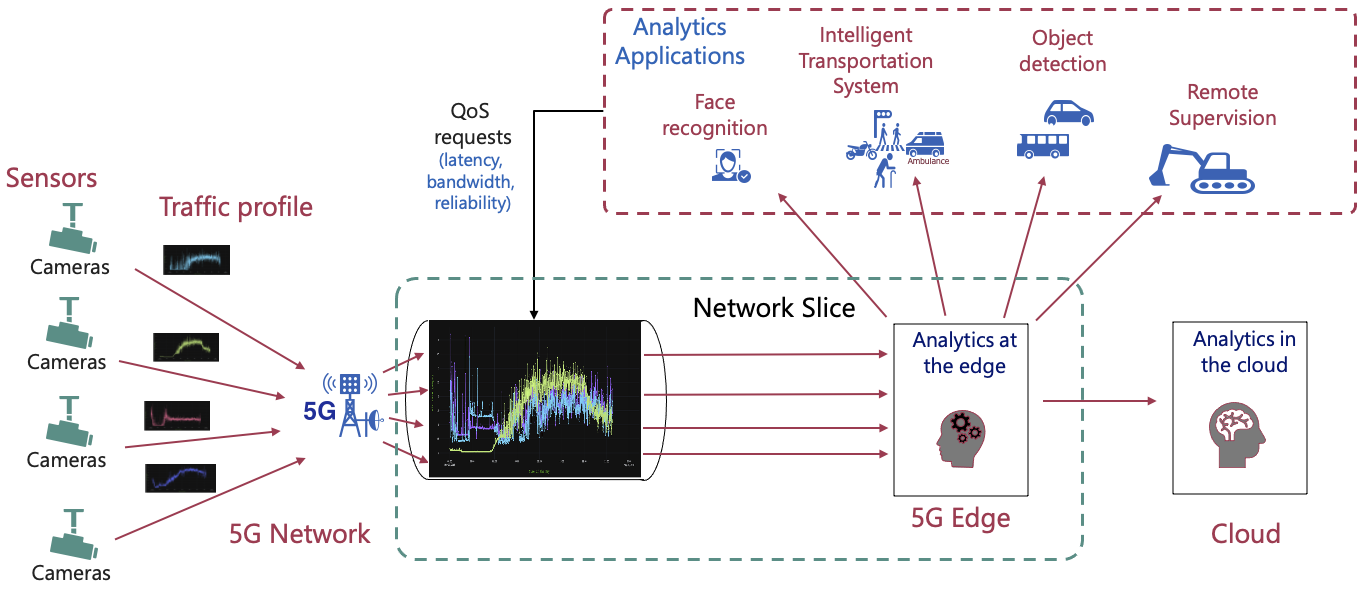}
\caption{System overview}
\label{fig:system-overview}
\end{figure*}

\section{System Overview}
\lSec{system-overview}
A high level system overview is shown in Fig. \ref{fig:system-overview}. Data streams from different sensors go over a 5G network into a network slice that has been set up by the 5G network for use by the analytics application, which can have tasks running at the edge. Depending on the network bandwidth needs of the sensors, the application can adapt and periodically issue updated QoS requests to the 5G network. As an illustration, consider the face recognition application, which recognizes faces seen in the video stream based on a pre-registered gallery of faces. The application monitors one or more video streams. These streams are made available to the application by the 5G network in a single network slice, whose QoS attributes are periodically updated by the face recognition application to ensure optimal use of the reserved network resources. For example, the bit rate of the video is low during the night when there is no movement of people in front of the camera. However, the bit rate is high during the day when people are moving in front of the camera. This time-varying, dynamic network traffic can be monitored by face recognition application and the bandwidth reserved at night can be way lower than the bandwidth reserved during the day. Different network bandwidth reservations during the night and day lead to significant network bandwidth savings, with almost no  effect on the accuracy of insights from the application.

\section{Cost function}
\label{sec:cost-function}
There are several forecasting methods used in various domains like weather prediction, hardware failure prediction, and exchange rate prediction in economics or finance.  In every case, it is desirable that the predicted value be as close as possible to the actual value. A cost function is typically minimized or maximized in order to achieve the desired prediction results, and cost functions are designed around traditional error metrics like Mean Absolute Error (MAE), Mean Absolute Percentage Error (MAPE), Mean Squared Error (MSE), Root Mean Squared Error (RMSE), Mean Direction Accuracy (MDA) and Adjusted Root Mean Squared Error (ARMSE) \cite{alternative-measure-of-forecasting-accuracy}.

The cost functions typically consider the magnitude or direction of the error, or some combination thereof.  For example, MAE is the average of the residuals or the absolute error (i.e. the difference between the predicted and actual value) and it considers only the magnitude of the error. It does not take into account the direction of the error (i.e. whether the predicted value is higher or lower than the actual value), or the frequency (i.e. how often the predicted value deviates from the actual, and in which direction). Similarly, MAPE also focuses on the percentage of the magnitude of the error, without considering direction or frequency. MSE is another error metric that is the average of the squared error and it measures the variance of the residuals. Again, like MAE, MSE also does not take into account the direction or frequency. RMSE is the square root of MSE and it measures the standard deviation of the residuals. MSE and RMSE penalize large prediction errors more, compared to MAE. Lower values of MAE, MSE and RMSE correspond to better predictions.  Unlike MAE, MSE and RMSE, MDA considers the direction of the predicted value, and it is a popular forecasting performance metric in economics and finance \cite{BERGMEIR2014132}. MDA captures the upward or downward trend but it is oblivious to the magnitude or frequency of the error. AMRSE tries to combine magnitude and direction, but it does not account for frequency. 

So, none of the known and popularly used cost functions simultaneously consider the magnitude, direction and frequency of the error. All three are important in our problem. The amount of actual network bandwidth usage (whether it is over or under the reserved network bandwidth), and how frequently the predicted usage is over or under the actual usage is important to minimize the operating costs of the application and avoid loss of accuracy of insights due to data loss. In our application, predicting usage that is less than the actual usage is critical because it can lead to undesirable loss in accuracy of insights produced by the analytics application.


We propose a new cost function (Equation \eqref{total-cost-eq}) that takes into account the magnitude, direction and frequency of the error. For each time unit $t$ in the total duration $T$, we check for undersubscription or oversubscription. The flag $F_u$ is 1 if there is undersubscription i.e. actual value ($A_t$), is less than reserved guaranteed bandwith (GBR) value $GBR_t$. The flag $F_o$ is 1 if there is oversubscription i.e. actual value ($A_t$) is greater than the GBR value $GBR_t$. Both $F_u$ and $F_o$ are zero otherwise, as shown in Equation \eqref{f-values}. Now, every time there is undersubscription, we penalize with a cost of $P_u$ and every time there is oversubscription, we penalize with a cost of $P_o$ and the penalty is applied to the actual amount of either undersubscription or oversubscription. The penalty for oversubscription is typically much higher than the penalty for undersubscription, since oversubscription directly adversely affects the accuracy of analytics. We want the oversubscription to be minimal, so that the accuracy of the analytics is not adversely affected. On the other hand, undersubscription leads to wastage of resources and higher operational costs due to excess reservation of network bandwidth. Therefore,  there is penalty for undersubscription too. However, this penalty is typically not as critical as the penalty for oversubscription. The actual penalty values can be configured depending on the relative importance of over and under subscription in the specific application.  The total cost is the sum of costs incurred for undersubscription and oversubscription, and lower the total cost, the better is the prediction.


\begin{equation}
\begin{split}
C_{total} & = \sum\limits_{t \in T} F_u\times{(GBR_t - A_t) \otimes P_u} \\
&+ {\sum\limits_{t \in T} F_o\times{(A_t - GBR_t) \otimes P_o}}
\end{split}
\label{total-cost-eq}
\end{equation}

{\small
\begin{equation}
    F_u =
    \begin{cases}
        1, &\text{if, } GBR_t > A_t \\
        0, &\text{otherwise}
    \end{cases}
    \text{and } 
    F_o =
    \begin{cases}
        1, &\text{if, } A_t > GBR_t \\
        0, &\text{otherwise}
    \end{cases}
    \label{f-values}
\end{equation}
}
\section{Prediction technique}
\label{sec:prediction-technique}
Different prediction techniques and models like ARMA, ARIMA, GARCH, Neural-network based, Linear regression, Support Vector Regression using linear and non-linear kernels e.g. RBF, Decision Tree based regression, Adaboost regression, etc. are used and applied to predict network traffic \cite{joshi2015review}. However, these prediction techniques do not account for the vastly different impact of oversubscription or undersubscription of network resource usage on the operating costs and accuracy of insights from the analytics application. Therefore, we propose a novel prediction technique, which analytics applications can use to dynamically self-optimize application's QoS requests to 5G networks. Our proposed prediction technique adequately takes into account the operating costs as well as the accuracy of insights from the analytics applications.

As mentioned in \rSec{cost-function}, oversubscription impacts the accuracy of the application and therefore is more critical. In our cost function, we assign a higher penalty for oversubscription. Our prediction technique aims to reduce oversubscription magnitude as well as frequency. We first design a simple prediction technique, which we call as the ``Max" prediction technique: we use the maximum network bandwidth that was actually utilized by the application in the previous interval as the predicted network bandwidth for the next interval. 
Now, with the ``Max" prediction technique, although we do consider and adjust to the upward or downward trend of the network bandwidth usage, it is still very coarse grained and may result in higher oversubscription or undersubscription, both in terms of magnitude as well as frequency. This is because, the network traffic may jump up or drop down at a very fast rate, and this is not easily captured quickly by the ``Max" prediction technique. Besides, blindly using the maximum may lead to significant undersubscription. To adjust to the rapid change in the rate of network bandwidth usage, avoid too much undersubscription, and in general to capture the upward or downward trend of the network bandwidth usage at a more fine-grained level, we propose next a slightly modified version of the ``Max" prediction technique, which we call as the  ``Modified-Max" technique.

In the ``Modified-Max" prediction technique, instead of just blindly using the maximum network bandwidth that was utilized by the application in the previous interval, we remember what was the predicted network bandwidth for the past $t$ time intervals. Now, for these $t$ time intervals, we see if there was an upward or downward trend of the network bandwidth usage. To be conservative and prevent future oversubscription, we consider that there is an upward trend even if we see that there is one occurrence of oversubscription in the past $t$ time intervals. Now, if there is an upward trend, then the mean of the magnitude of oversubscription above the predicted value of all occurrences of oversubscription is calculated. The highest magnitude or the maximum of the network bandwidth utilized is then chosen as the baseline for the next prediction and the calculated mean is \textit{added} to this baseline, which adjusts or modifies the maximum and this ``modified maximum" is then chosen as the next network bandwidth prediction.

Along with considering the upward trend, ``Modified-Max" prediction technique also takes into account the downward trend i.e. when there is no oversubscription observed in the past $t$ time intervals. When there is a downward trend, then the mean of the magnitude of undersubscription below the predicted value of all occurrences of undersubscription is calculated. This mean is then \textit{subtracted} from the baseline, which is the highest magnitude or the maximum of the network bandwidth utilized by the application, and this new ``modified maximum" is then chosen as the prediction for the next interval. Thus, when there is an upward trend, ``Modified-Max" technique raises the predicted network bandwidth for next interval above the baseline, while for downward trend, it lowers the predicted network bandwidth below the baseline, thereby, capturing the rapid change in the rate of network bandwidth usage as well as capturing the upward or downward trend in a much finer granularity. Our ``Modified-Max" prediction technique is able to keep the oversubscription as well as undersubscription in check, while giving priority to prevent oversubscription more than undersubscription.

\section{Experiments and Results}
\lSec{experiments-results}

\begin{figure}[t]
\begin{subfigure}[]{0.99\linewidth}
\centering
    \includegraphics[height = 1.5 in]{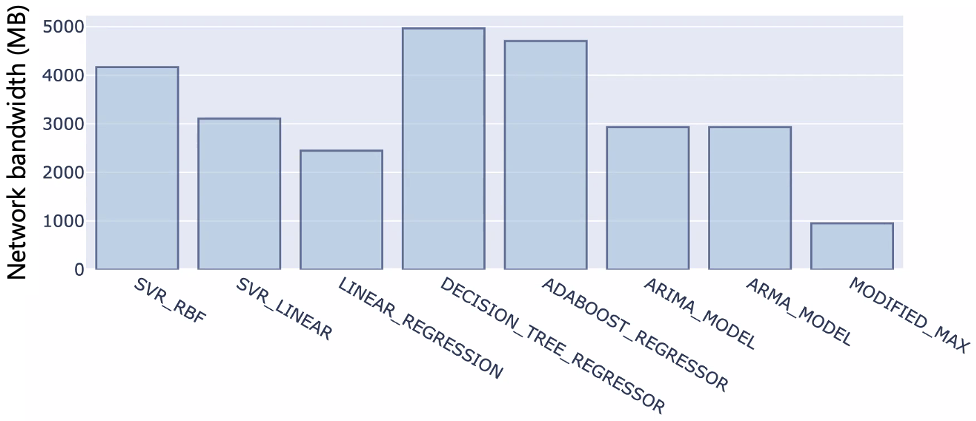}
    \caption{Magnitude of oversubscription}
\end{subfigure}
\begin{subfigure}[]{0.99\linewidth}
\centering
    \includegraphics[height = 1.5 in]{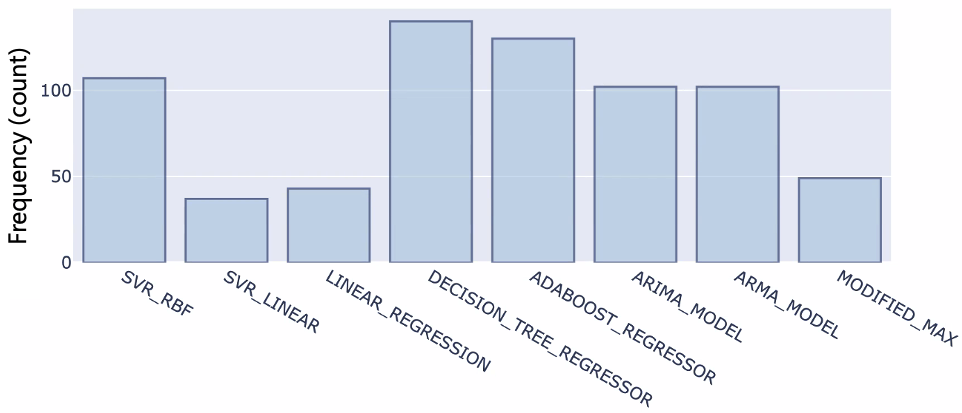}
    \caption{Frequency of oversubscription}
\end{subfigure}
\begin{subfigure}[]{0.99\linewidth}
\centering
    \includegraphics[height = 1.5 in]{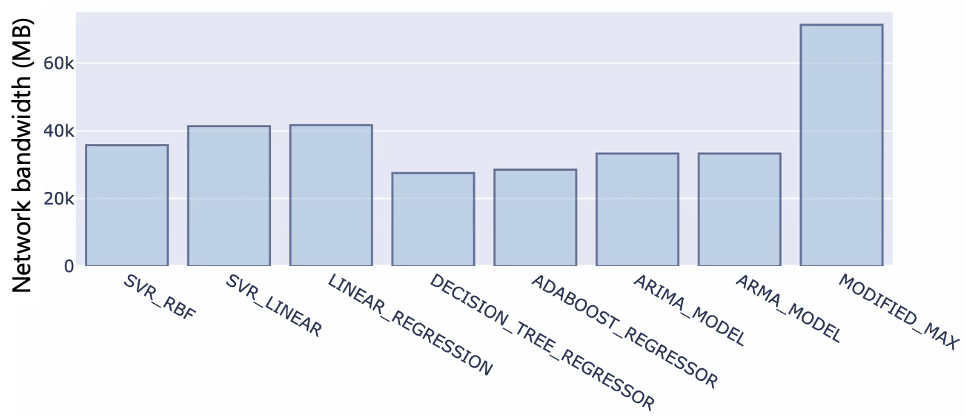}
    \caption{Magnitude of undersubscription}
\end{subfigure}
\begin{subfigure}[]{0.99\linewidth}
\centering
    \includegraphics[height = 1.5 in]{images/frequency-oversubscription.png}
    \caption{Frequency of undersubscription}
\end{subfigure}
\caption{Magnitude and frequency for different techniques}
\label{magnitude}
\end{figure}

We report our experience with a real-world deployment of a video surveillance application that monitors 17 cameras in a busy airport.  These cameras are placed indoors and outdoors, they experience different lighting conditions, and they monitor different locations with time-varying people traffic.  Video streams from these cameras are sent over a private 5G network for remote processing. Our application uses a single network slice to access all the 17 video streams.

We use \textit{Axis Q1615} cameras with H.264 encoding for the video streams. As the environment changes, the network traffic profiles (i.e. bit rates from these individual cameras) change. The bit rate varies for an individual camera. The bit rate also varies across different cameras due to the time-varying density of people traffic (network traffic profiles for some of the cameras is shown in Fig. \ref{network-traffic-profiles}). For our video analytics application, it is important to predict the network bandwidth requirement for each camera, aggregate the predictions, and issue a single QoS request to the private 5G network to change the guaranteed bandwidth attribute of the network slice. 


Fig. \ref{magnitude} compares various prediction techniques in terms of the magnitude and frequency of oversubscription,  and the magnitude of undersubscription. We observe that the magnitude of oversubscription for the proposed  ``Modified-Max"  technique is much less than the others. We also observe that the frequency of oversubscription is quite low, compared to others. This shows that our technique is very effective in preventing oversubscription (data loss), thereby ensuring high analytics application accuracy.

We also observe that the magnitude of undersubscription for the proposed ``Modified-Max"  technique is higher than the other prediction techniques (because we use ``Max" value from previous interval as our baseline). However, when compare to a static, one-time reservation, our technique saves 34\% of network bandwidth. We report compared to static, one-time, reservation rather than other prediction techniques because other prediction techniques have a high magnitude of oversubscription, which is not acceptable for the application. So, if zero \textit{oversubscription} is desired, then the application has to reserve the worst case network bandwidth and incur the associated high network resource reservation costs. 

\begin{figure}[ht]
\centering
\includegraphics[width=0.99\linewidth]{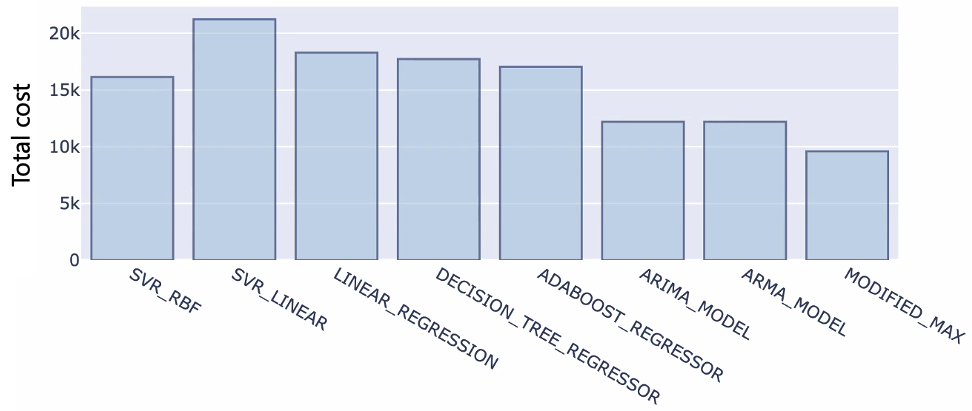}
\caption{Total cost for different prediction techniques}
\label{total-cost}
\end{figure}

Fig. \ref{total-cost} compares the proposed cost function values for the different techniques. We used the same oversubscription and undersubscription penalty for all the techniques. The exact values used were 0.1 for undersubscription penalty ($P_u$) and 30 for oversubscription penalty ($P_o$). We observe that our ``Modified-Max" prediction technique has the lowest cost and it outperforms all the other prediction techniques.
\section{Conclusion}
\lSec{conclusion}
Network traffic profile from sensors e.g. video cameras, vary depending on environmental conditions, and analytics applications receiving such time-varying, dynamic sensor data streams over 5G networks, have to decide and issue QoS requests to 5G networks for network slices, since 5G networks are unaware of the application requirements. Static, one-time, issuance of QoS request by application, leads to tremendous amount of wastage of network resources, which increases operational cost significantly. To this end, in this paper, we propose a novel QoS request prediction technique, specifically, network bandwidth prediction technique, and we also present a new cost function, which takes into account the magnitude, direction and frequency of error i.e. difference between the actual used and reserved network bandwidth. Our proposed technique empowers applications to dynamically adapt and self-optimize QoS requests to 5G networks in order to reduce network bandwidth wastage, and thereby operational cost, with minimal effect on accuracy. With real-world deployment of video analytics application over 17 cameras, we show that our technique is able to save 34\% of network bandwidth, with minimal data loss, when compared to a static, one-time reservation and it also outperforms other traditional forecasting methods.
\bibliographystyle{IEEEtran}


\end{document}